# 4MOST: Project overview and information for the First Call for Proposals

Roelof S. de Jong[1,a]
Oscar Agertz[2]
Alex Agudo Berbel[3]
James Aird[4]
David A. Alexander[5]
Anish Amarsi[6]
Friedrich Anders[1]
Rene Andrae[7]
Behzad Ansarinejad[5]
Wolfgang Ansorge[8]
Pierre Antilogus[9]
Heiko Anwand-Heerwart[10]
Anke Arentsen[1]
Anna Arnadottir[2]
Martin Asplund[6]
Matt Auger[4]
Nicolas Azais[1,11]
Dietrich Baade[12]
Gabriella Baker[13]
Sufyan Baker[13]
Eduardo Balbinot[14]
Ivan K. Baldry[15]
Manda Banerji[4]
Samuel Barden[1]
Paul Barklem[16]
Eléonore Barthélémy-Mazot[17]
Chiara Battistini[18]
Svend Bauer[1]
Cameron P. M. Bell[1]
Olga Bellido-Tirado[1]
Sabine Bellstedt[19]
Vasily Belokurov[4]
Thomas Bensby[2]
Maria Bergemann[7]
Joachim M. Bestenlehner[20]
Richard Bielby[5]
Maciej Bilicki[21]
Chris Blake[22]
Joss Bland-Hawthorn[23]
Corrado Boeche[24]
Wilfried Boland[25,21]
Thomas Boller[3]
Sebastien Bongard[9]
Angela Bongiorno[26]
Piercarlo Bonifacio[27]
Didier Boudon[28]
David Brooks[29]
Michael J. I. Brown[30]
Rebecca Brown[13]
Marcus Brüggen[31]
Joar Brynnel[1]
Jurek Brzeski[13]
Thomas Buchert[28]
Peter Buschkamp[18]
Elisabetta Caffau[27]
Patrick Caillier[28]
Jonathan Carrick[32]
Luca Casagrande[6]
Scott Case[13]
Andrew Casey[30]
Isabella Cesarini[1]
Gabriele Cescutti[33]
Diane Chapuis[28]
Cristina Chiappini[1]
Michael Childress[34]
Norbert Christlieb[18]
Ross Church[2]
Maria-Rosa L. Cioni[1]
Michelle Cluver[22]
Matthew Colless[6]
Thomas Collett[35]
Johan Comparat[3]
Andrew Cooper[5]
Warrick Couch[36,22]
Frederic Courbin[37]
Scott Croom[23]
Darren Croton[22]
Eric Daguisé[28]
Gavin Dalton[38]
Luke J. M. Davies[19]
Tamara Davis[39]
Patrick de Laverny[40]
Alis Deason[5]
Frank Dionies[1]
Karen Disseau[28]
Peter Doel[29]
Daniel Döscher[1]
Simon P. Driver[19]
Tom Dwelly[3]
Dominique Eckert[3]
Alastair Edge[5]
Bengt Edvardsson[16]
Dalal El Youssoufi[1]
Ahmed Elhaddad[18]
Harry Enke[1]
Ghazaleh Erfanianfar[3]
Tony Farrell[13]
Thomas Fechner[1]
Carmen Feiz[18]
Sofia Feltzing[2]
Ignacio Ferreras[29]
Dietrich Feuerstein[1]
Diane Feuillet[7]
Alexis Finoguenov[3,41]
Dominic Ford[2]
Sotiria Fotopoulou[5]
Morgan Fouesneau[7]
Carlos Frenk[5]
Steffen Frey[1]
Wolfgang Gaessler[7]
Stephan Geier[42]
Nicola Gentile Fusillo[43]
Ortwin Gerhard[3]
Tommaso Giannantonio[4]
Domenico Giannone[1]
Brad Gibson[44]
Peter Gillingham[13]
Carlos González-Fernández[4]
Eduardo Gonzalez-Solares[4]
Stefan Gottloeber[1]
Andrew Gould[45,7]
Eva K. Grebel[24]
Alain Gueguen[3]
Guillaume Guiglion[1]
Martin Haehnelt[4]
Thomas Hahn[1]
Camilla J. Hansen[7,46]
Henrik Hartman[2]
Katja Hauptner[10]
Keith Hawkins[4]
Dionne Haynes[1]
Roger Haynes[1]
Ulrike Heiter[16]
Amina Helmi[14]
Cesar Hernandez Aguayo[5]
Paul Hewett[4]
Samuel Hinton[39]
David Hobbs[2]
Sebastian Hoenig[34]
David Hofman[17]
Isobel Hook[32]
Joshua Hopgood[12]
Andrew Hopkins[13]
Anna Hourihane[4]
Louise Howes[2]
Cullan Howlett[19]
Tristan Huet[1]
Mike Irwin[4]
Olaf Iwert[12]
Pascale Jablonka[37]
Thomas Jahn[1]
Knud Jahnke[7]
Aurélien Jarno[28]
Shoko Jin[14]
Paula Jofre[4]
Diana Johl[1]
Damien Jones[47]
Henrik Jönsson[2]
Carola Jordan[7]
Iva Karovicova[18]
Arman Khalatyan[1]
Andreas Kelz[1]
Robert Kennicutt[4]
David King[4]
Francisco Kitaura[48]
Jochen Klar[1]
Urs Klauser[13]
Jean-Paul Kneib[37]
Andreas Koch[24]
Sergey Koposov[4]
Georges Kordopatis[40]
Andreas Korn[16]
Johan Kosmalski[12,28]
Rubina Kotak[49,50]
Mikhail Kovalev[7]
Kathryn Kreckel[7]
Yevgen Kripak[13]

Mirko Krumpe[1]
Koen Kuijken[21]
Andrea Kunder[1]
Iryna Kushniruk[2]
Man I Lam[1]
Georg Lamer[1]
Florence Laurent[28]
Jon Lawrence[13]
Michael Lehmitz[7]
Bertrand Lemasle[24]
James Lewis[4]
Baojiu Li[5]
Chris Lidman[36,6]
Karin Lind[16]
Jochen Liske[31]
Jean-Louis Lizon[12]
Jon Loveday[51]
Hans-Günter Ludwig[18]
Richard M. McDermid[52]
Kate Maguire[49]
Vincenzo Mainieri[12]
Slavko Mali[13]
Holger Mandel[18]
Kaisey Mandel[4]
Liz Mannering[36,19]
Sarah Martell[53]
David Martinez Delgado[24]
Gal Matijevic[1]
Helen McGregor[13]
Richard McMahon[4]
Paul McMillan[2]
Olga Mena[54]
Andrea Merloni[3]
Martin J. Meyer[19]
Christophe Michel[17]
Genoveva Micheva[1]
Jean-Emmanuel Migniau[28]
Ivan Minchev[1]
Giacomo Monari[1]
Rolf Muller[13]
David Murphy[4]
Daniel Muthukrishna[4]
Kirpal Nandra[3]
Ramon Navarro[55]
Melissa Ness[7]
Vijay Nichani[13]
Robert Nichol[35]
Harald Nicklas[10]
Florian Niederhofer[1]
Peder Norberg[5]
Danail Obreschkow[19]
Seb Oliver[51]
Matt Owers[52]
Naveen Pai[13]
Sergei Pankratow[1]
David Parkinson[39]
Jens Paschke[1]
Robert Paterson[13]
Arlette Pecontal[28]
Ian Parry[4]
Dan Phillips[1]
Annalisa Pillepich[7]
Laurent Pinard[17]
Jeff Pirard[12]
Nikolai Piskunov[16]
Volker Plank[1]
Dennis Plüschke[1]
Estelle Pons[4]
Paola Popesso[56]
Chris Power[19]
Johan Pragt[55]
Alexander Pramskiy[18]
Dan Pryer[51]
Marco Quattri[12]
Anna Barbara de Andrade Queiroz[1]
Andreas Quirrenbach[18]
Swara Rahurkar[1]
Anand Raichoor[37]
Sofia Ramstedt[16]
Arne Rau[3]
Alejandra Recio-Blanco[40]
Roland Reiss[12]
Florent Renaud[2]
Yves Revaz[37]
Petra Rhode[10]
Johan Richard[28]
Amon David Richter[10]
Hans-Walter Rix[7]
Aaron S. G. Robotham[19]
Ronald Roelfsema[57,55]
Martino Romaniello[12]
David Rosario[5]
Florian Rothmaier[18]
Boudewijn Roukema[58,28]
Gregory Ruchti[2]
Gero Rupprecht[12]
Jan Rybizki[7]
Nils Ryde[2]
Andre Saar[1]
Elaine Sadler[23]
Martin Sahlén[16]
Mara Salvato[3]
Benoit Sassolas[17]
Will Saunders[13]
Allar Saviauk[1]
Luca Sbordone[59]
Thomas Schmidt[1]
Olivier Schnurr[1,60]
Ralf-Dieter Scholz[1]
Axel Schwope[1]
Walter Seifert[18]
Tom Shanks[5]
Andrew Sheinis[36,61]
Tihomir Sivov[1]
Ása Skúladóttir[7]
Stephen Smartt[49]
Scott Smedley[13]
Greg Smith[1]
Robert Smith[51]
Jenny Sorce[28,1]
Lee Spitler[52]
Else Starkenburg[1]
Matthias Steinmetz[1]
Ingo Stilz[18]
Jesper Storm[1]
Mark Sullivan[34]
William Sutherland[62]
Elizabeth Swann[35]
Amélie Tamone[37]
Edward N. Taylor[22]
Julien Teillon[17]
Elmo Tempel[63,1]
Rik ter Horst[55]
Wing-Fai Thi[3]
Eline Tolstoy[14]
Scott Trager[14]
Gregor Traven[2]
Pier-Emmanuel Tremblay[43]
Laurence Tresse[28]
Marica Valentini[1]
Rien van de Weygaert[14]
Mario van den Ancker[12]
Jovan Veljanoski[14]
Sudharshan Venkatesan[13]
Lukas Wagner[1]
Karl Wagner[18]
C. Jakob Walcher[1]
Lew Waller[13]
Nicholas Walton[4]
Lingyu Wang[57,14]
Roland Winkler[1]
Lutz Wisotzki[1]
C. Clare Worley[4]
Gabor Worseck[42]
Maosheng Xiang[7]
Wenli Xu[64]
David Yong[6]
Cheng Zhao[37]
Jessica Zheng[13]
Florian Zscheyge[1]
Daniel Zucker[52]

[1] Leibniz-Institut für Astrophysik Potsdam (AIP), Germany
[2] Lund Observatory, Lund University, Sweden
[3] Max-Planck-Institut für extraterrestrische Physik, Garching, Germany
[4] Institute of Astronomy, University of Cambridge, UK
[5] Department of Physics, Durham University, UK
[6] Research School of Astronomy & Astrophysics, Australian National University, Canberra, Australia

[7] Max-Planck-Institut für Astronomie, Heidelberg, Germany
[8] RAMS-CON, Assling, Germany
[9] Laboratoire de physique nucléaire et de hautes énergies, Paris, France
[10] Institut für Astrophysik, Georg-August Universität Göttingen, Germany
[11] IRIDESCENCE, Paris, France
[12] ESO
[13] Australian Astronomical Optics — Macquarie, Sydney, Australia
[14] Kapteyn Instituut, Rijksuniversiteit Groningen, the Netherlands
[15] Astrophysics Research Institute, Liverpool John Moores University, UK
[16] Department of Physics and Astronomy, Uppsala universitet, Sweden
[17] Laboratoire des Matériaux Avancés, Lyon, France
[18] Zentrum für Astronomie der Universität Heidelberg/Landessternwarte, Germany
[19] International Centre for Radio Astronomy Research/University of Western Australia, Perth, Australia
[20] Physics and Astronomy, University of Sheffield, UK
[21] Sterrewacht Leiden, Universiteit Leiden, the Netherlands
[22] Centre for Astrophysics and Supercomputing, Swinburne University of Technology, Hawthorn, Australia
[23] Sydney Institute for Astronomy, University of Sydney, Australia
[24] Zentrum für Astronomie der Universität Heidelberg/Astronomisches Rechen-Institut, Germany
[25] Nederlandse Onderzoekschool Voor Astronomie (NOVA), Leiden, the Netherlands
[26] Osservatorio Astronomico di Roma, INAF, Italy
[27] GEPI, Observatoire de Paris, Université PSL, CNRS, France
[28] Centre de Recherche Astrophysique de Lyon, France
[29] Department of Physics and Astronomy, University College London, UK
[30] School of Physics and Astronomy, Monash University, Melbourne, Australia
[31] Hamburger Sternwarte, Universität Hamburg, Germany
[32] Physics Department, Lancaster University, UK
[33] Osservatorio Astronomico di Trieste, INAF, Italy
[34] School of Physics and Astronomy, University of Southampton, UK
[35] Institute of Cosmology and Gravitation, University of Portsmouth, UK
[36] Australian Astronomical Observatory, Sydney, Australia
[37] Laboratoire d'astrophysique, École Polytechnique Fédérale de Lausanne, Switzerland
[38] Department of Physics, University of Oxford, UK
[39] School of Mathematics and Physics, University of Queensland, Brisbane, Australia
[40] Observatoire de la Côte d'Azur, Nice, France
[41] University of Helsinki, Finland
[42] Institut für Physik und Astronomie, Universität Potsdam, Germany
[43] Department of Physics, University of Warwick, UK
[44] E. A. Milne Centre for Astrophysics, University of Hull, UK
[45] Ohio State University, Columbus, USA
[46] Dark Cosmology Centre, Københavns Universitet, Denmark
[47] Prime Optics, Eumundi, Queensland, Australia
[48] Instituto de Astrofísica de Canarias, La Laguna, Tenerife, Spain
[49] School of Mathematics and Physics, Queen's University Belfast, UK
[50] University of Turku, Finland
[51] University of Sussex, Brighton, UK
[52] Department of Physics and Astronomy, Macquarie University, Sydney, Australia
[53] School of Physics, University of New South Wales, Sydney, Australia
[54] Instituto de Física Corpuscular, Universidad de Valencia, Spain
[55] Nederlandse Onderzoekschool Voor Astronomie (NOVA), Dwingeloo, the Netherlands
[56] Physics Department, Technische Universität München, Germany
[57] Netherlands Institute for Space Research (SRON), Groningen, the Netherlands
[58] Torun Centre for Astronomy (TCfA), Nicolaus Copernicus University, Poland
[59] Pontificia Universidad Católica de Chile, Santiago, Chile
[60] Cherenkov Telescope Array Observatory, Bologna, Italy
[61] CFHT, Kamuela, Hawaii, USA
[62] School of Physics and Astronomy, Queen Mary University of London, UK
[63] Tartu Observatory, University of Tartu, Estonia
[64] XU-OSE, Heidelberg, Germany

**We introduce the 4-metre Multi-Object Spectroscopic Telescope (4MOST), a new high-multiplex, wide-field spectroscopic survey facility under development for the four-metre-class Visible and Infrared Survey Telescope for Astronomy (VISTA) at Paranal. Its key specifications are: a large field of view (FoV) of 4.2 square degrees and a high multiplex capability, with 1624 fibres feeding two low-resolution spectrographs ($R = \lambda/\Delta\lambda \sim 6500$), and 812 fibres transferring light to the high-resolution spectrograph ($R \sim 20\,000$). After a description of the instrument and its expected performance, a short overview is given of its operational scheme and planned 4MOST Consortium science; these aspects are covered in more detail in other articles in this edition of The Messenger. Finally, the processes, schedules, and policies concerning the selection of ESO Community Surveys are presented, commencing with a singular opportunity to submit Letters of Intent for Public Surveys during the first five years of 4MOST operations.**

4MOST is being developed to address a broad range of pressing scientific questions in the fields of Galactic archaeology, high-energy astrophysics, galaxy evolution and cosmology. Its design allows tens of millions of spectra to be obtained via five-year surveys, even for targets distributed over a significant fraction of the sky. While many science cases can be addressed with 4MOST, its primary purpose is to provide the spectroscopic complements to large-area surveys coming from key European space missions like eROSITA and the ESA Gaia, Euclid and PLATO missions, as well as from ground-based facilities like VISTA, the VLT Survey Telescope (VST), the Dark Energy Survey (DES), the Large Synoptic Survey Telescope (LSST) and the Square Kilometre Array (SKA).

Multiple science cases must be carried out simultaneously in order to efficiently fill all the fibres in a high multiplex instrument like 4MOST. This necessitates effective coordination between different science teams. To enable this, the 4MOST Consortium will perform Public Surveys using 70% of the available fibre-hours in the first five years of operation.

These Public Surveys are Guaranteed Time Observations (GTO) that the Consortium receives in return for building the facility and for supporting ESO in the operation of 4MOST. Public Surveys of the ESO and the Chilean host country communities will fill the other 30% of available fibre-hours in the first five years of operation. These surveys will be chosen by a one-time, competitive, peer-reviewed selection process, similarly to other ESO Calls for Public Surveys. Here, a fibre-hour is defined as one hour of observing time, including overheads, with one fibre; hence 4MOST offers 2436 fibre-hours every hour that it is observing.

Following this overview, which contains information on instrument performance and on the procedures associated with the use of 4MOST by the community, this issue of The Messenger includes additional articles on the 4MOST science operations model, the survey plan of the 4MOST Consortium, and a description of the ten Public Surveys that the Consortium intends to carry out. Together these articles are intended to prepare the ESO community for the proposal process that will commence in the second half of 2019. The process will start with a one-off opportunity for the submission of Letters of Intent to apply for Public Surveys to be executed during the first five years of 4MOST operation.

## Organisation

The 4MOST project is organised along three branches:
1. Instrument — responsible for the development, construction, and commissioning of the instrument hardware and associated software;
2. Operations — for the planning, data reduction, archiving, and publishing of the observations including the associated data-flow;
3. Science — the branch that develops the different Surveys and is responsible for science analysis and publication.

The instrument and operations branches are mainly performed by the 4MOST Consortium and are jointly called the 4MOST Facility.

Table 1. 4MOST key instrument specifications.

| Instrument parameter | Design value |
|---|---|
| Field of view (hexagon) | ~ 4.2 square degrees (Ø = 2.6 degrees) |
| Accessible sky (zenith angle < 55 degrees) | > 30 000 square degrees |
| Expected on-target fibre-hours per year | LRS: > 3 200 000 h $yr^{-1}$, HRS > 1 600 000 h $yr^{-1}$ |
| Multiplex fibre positioner | 2436 |
| Low-Resolution Spectrographs LRS (× 2) | |
| Resolution | $<R>$ = 6500 |
| Number of fibres | 812 fibres |
| Passband | 3700–9500 Å |
| Velocity accuracy | < 1 km $s^{-1}$ |
| Mean sensitivity 6 × 20 min, mean seeing, new moon, S/N = 10 $Å^{-1}$ (AB-magnitude) | 4000 Å: 20.2, 5000 Å: 20.4, 6000 Å: 20.4, 7000 Å: 20.2, 8000 Å: 20.2, 9000 Å: 19.8 |
| High-Resolution Spectrograph HRS (× 1) | |
| Resolution | $<R>$ = 20 000 |
| Number of fibres | 812 fibres |
| Passband | 3926–4355, 5160–5730, 6100–6790 Å |
| Velocity accuracy | < 1 km $s^{-1}$ |
| Mean sensitivity 6 × 20 min, mean seeing, 80% moon, S/N = 100 $Å^{-1}$ (AB-magnitude) | 4200 Å: 15.7, 5400 Å: 15.8, 6500 Å: 15.8 |
| Smallest target separation | 15 arcseconds on any side |
| # of fibres in random Ø = 2 arcminute circle | ≥ 3 |
| Fibre diameter | Ø = 1.45 arcseconds |

The instrument is under construction at a number of Consortium institutes, coordinated by the 4MOST Project Office located at the Leibniz-Institut für Astrophysik Potsdam (AIP). Once the subsystems are finished at the different institutes, they will all be transported to Potsdam and extensively tested there as a full system before being shipped to Paranal. At Paranal the 4MOST instrument will be installed, tested, and commissioned on the VISTA telescope.

The operations branch is led by the Operations Development Group, consisting of the leads of the different subsystems and working groups involved in observation planning and data-flow. It also contains the 4MOST Helpdesk activities.

The science programme is organised into several surveys. The members of the survey teams are spread over all participating institutes and each team is led by one or more Survey Principal Investigators (Survey PIs). Coordination between all participating surveys is performed by the Science Coordination Board (SCB), consisting of all Survey PIs. The science branch is overseen by two Project Scientists, one for Galactic and one for extragalactic science, who have both a science guidance and a managerial role.

## Instrument

The 4MOST instrument design was driven by the science requirements of its key Consortium Surveys. Within a 2-hour observation 4MOST has the sensitivity to obtain redshifts of $r$ = 22.5 magnitudes (AB) galaxies and active galactic nuclei (AGN), radial velocities of any Gaia source ($G$ < 20.5 magnitudes [Vega]), stellar parameters and selected key elemental abundances with accuracy better than 0.15 dex of $G$ < 18-magnitude stars, and abundances of up to 15 elements of $G$ < 15.5-magnitude stars. Furthermore, in a five-year survey 4MOST can cover > 17 000 square degrees at least twice and obtain spectra of more than 20 million sources with a resolution of $R$ ~ 6500 and more than three million spectra with a resolution of $R$ ~ 20 000 for the typical science cases proposed. The main instrument parameters enabling these science requirements are summarised in Table 1.

Figure 1 provides an overview of the main instrument subsystems. A new Wide Field Corrector (WFC) equipped with an Atmospheric Dispersion Compensator (ADC) that provides corrections to a 55-degree zenith angle distance creates a focal surface with a 2.6-degree diameter. Two Acquisition and Guiding (A&G) cameras ensure correct pointing, while four

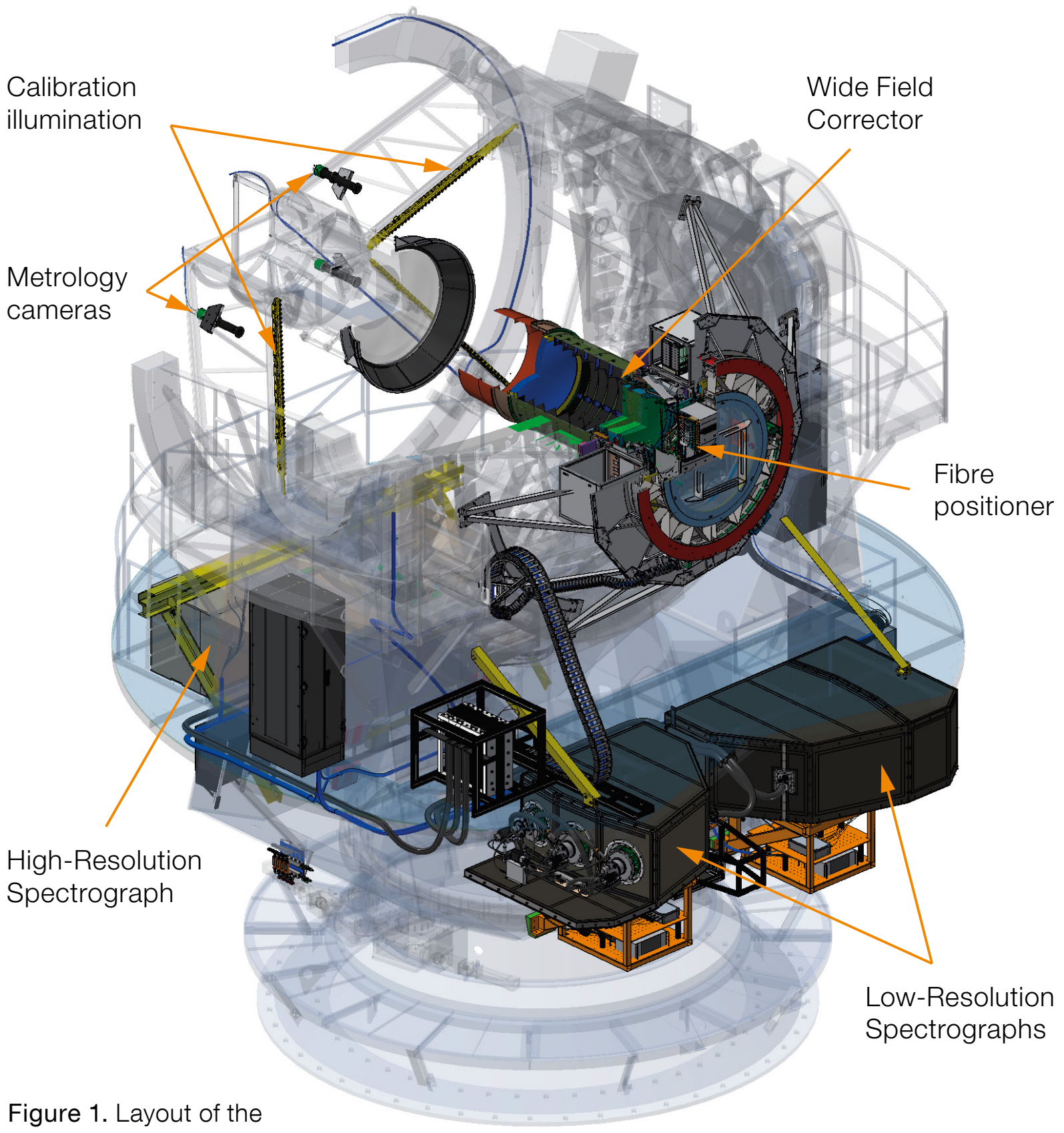

Figure 1. Layout of the different subsystems of 4MOST on the VISTA telescope.

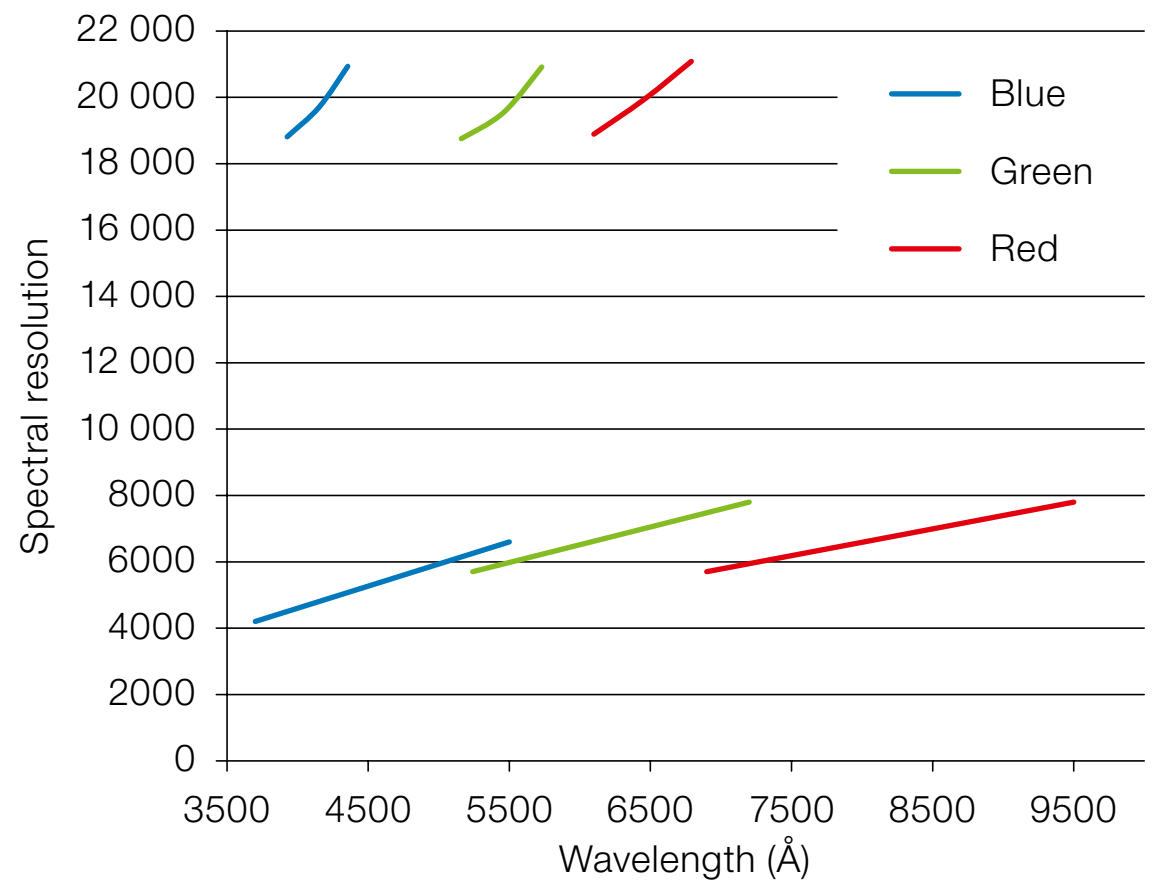

Figure 2. Spectral resolution in the three channels of the 4MOST High-Resolution (HRS, upper lines) and Low-Resolution Spectrographs (LRS; lower lines).

Wave Front Sensing (WFS) cameras steer the active optics system of the telescope.

The AESOP fibre positioning system based on the tilting spine principle can, within 2 minutes, simultaneously position all of the 2436 science fibres that are arranged in a hexagonally shaped grid at the focal surface. The accuracy of fibre positioning is expected to be better than 0.2 arcseconds thanks to a four-camera metrology system observing the fibre tips back-illuminated from the spectrograph. The tilting spine positioner has the advantage that each fibre has a large patrol area; each target in the science field of view can be reached by at least three fibres that go to one of the Low-Resolution Spectrographs (LRS) and one or two fibres that go to the High-Resolution Spectrograph (HRS). This ensures a high allocation efficiency of the fibres to targets, even when targets are clustered.

Each spectrograph accepts 812 science fibres and six simultaneous calibration fibres attached to either end of the spectrograph entrance slit. The covered wavelength range and resolution of the LRS and HRS spectrographs are as listed in Table 1 and depicted in Figure 2. Each type of spectrograph has three channels in fixed configurations covering three wavelength bands, and is thermally invariant and insulated (HRS) or temperature controlled (LRS) for stability. Each channel is equipped with a 6 k × 6 k CCD detector with low read noise (< 2.3 electrons per read) and with high, broadband quantum efficiency. The spectra are sampled with about three pixels per resolution element.

A calibration system equipped with a continuum source, a Fabry-Perot etalon, and ThAr lamps can feed light through the telescope plus science fibres combination and also directly through the simultaneous calibration fibres into the spectrograph slit to ensure accurate wavelength calibration. This will ensure that we can typically reach better than 1 km $s^{-1}$ accuracy on stellar radial velocities. The expected sensitivity is depicted in Figure 3. The estimated observing overheads are currently conservatively estimated to be 3.5 minutes per repointing of the telescope and 4.4 minutes per science exposure for repositioning of the fibres, obtaining attached calibration frames, and performing detector readout. We aim to reduce these overhead numbers in the future by executing more exposure setup activities in parallel and by reducing the number of attached night-time calibration exposures once we have established the stability and calibration reproducibility of the full system.

## Operations

The 4MOST operations scheme differs from other ESO instrument operations in that it allows many different science cases to be scheduled simultaneously during one observation. To accommodate the range of exposure times required for different targets, the same part of the sky will be observed with multiple exposures and visits. Objects that require longer exposures will be exposed several times until their stacked spectra reach the required signal-to-noise. 4MOST operations also differ from the standard ESO scheme in that the 4MOST Consortium plays a primary role in planning the observations (Phase 2) and in reducing, analysing and publishing the data (Phase 3).

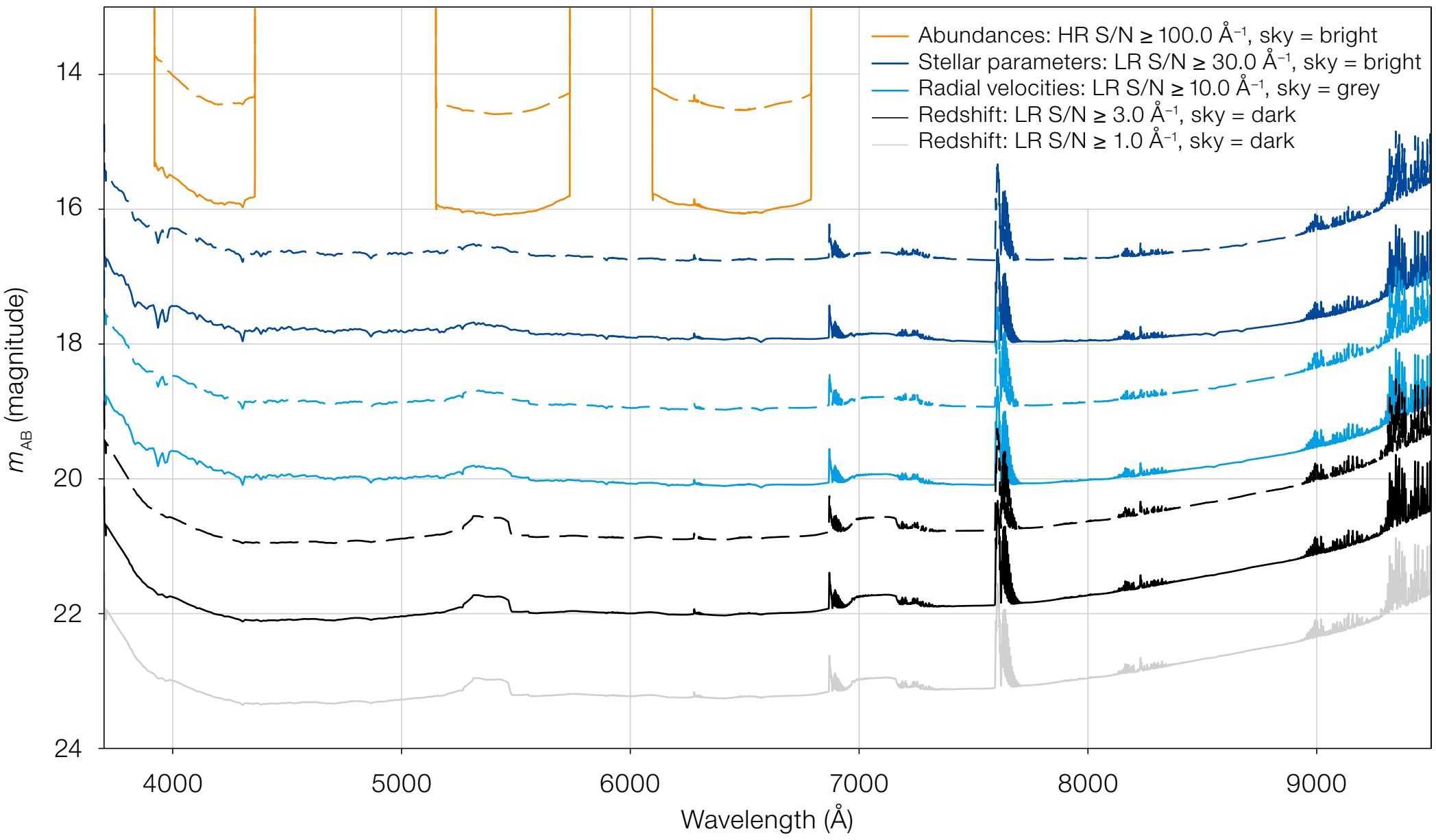

Figure 3. The expected 4MOST point-source sensitivities for the signal-to-noise levels and lunar conditions indicated in the legend. The solid lines are for a total exposure time of 120 minutes, whereas the dashed lines are the limits for 20-minute exposures. The approximate conversion to signal-to-noise per pixel is obtained by dividing the HRS values by 3.3 and the LRS values by 1.7. For clarity, sky emission lines are removed — this mostly affects results redward of 7000 Å. Mean (not median) seeing conditions, airmass values, fibre quality and positioning errors, etc., are used, in order to ensure that this plot is representative for an entire 4MOST survey, not just for the optimal conditions. Typical science cases for obtaining detailed elemental abundances of stars (orange), stellar parameters and some elemental abundances (dark blue), stellar radial velocities (light blue), and galaxy and AGN redshifts (black: 90% complete, grey: 50% complete) are shown.

These Consortium activities are closely monitored by ESO to ensure uniform progress and data quality for all surveys. The details of 4MOST operations are described in the accompanying article in this edition of The Messenger (Walcher et al., p. 12).

## Science

The 4MOST science programme formulated by the Consortium has been organised into the ten surveys listed in Table 2. There are five surveys centred on stellar objects to perform Galactic archaeology of different components of the Milky Way and the Magellanic Clouds, with the goal of understanding their current structure and their assembly history. There are four surveys of extragalactic objects aiming to characterise cosmological parameters, the nature of dark energy and dark matter, and the formation history of galaxies and black holes. Finally, there is a survey dedicated to time domain discoveries, mainly in synergy with the LSST facility where supernova transients and quasar luminosity variations will be complemented with spectroscopic observations.

For most of these surveys, millions of spectra will be obtained, having a huge legacy value for the community and creating an enormous potential for serendipitous discoveries. Being the only facility in the south with such a large field of view and multiplex capability creates numerous unique opportunities for 4MOST. Of special interest are synergies with new southern hemisphere facilities under construction such as LSST, SKA, and ESO's ELT. The southern sky is of particular interest for Galactic archaeology, with good access to the Milky Way bulge and the Magellanic Clouds. For this science, the $R$ ~ 20 000 of the HRS enables accurate abundance measurements of many elements; the $R$ ~ 6500 LRS spectra also have higher spectral resolution and better sampling of the spectral resolution elements than similar high-multiplex, wide-field facilities, thereby allowing better stellar elemental abundance determinations. 4MOST provides an unprecedentedly large volume coverage of all Galactic components, thereby expanding on the legacy of the ESA Gaia mission.

This Messenger edition contains sufficiently detailed descriptions of the Consortium Surveys and the overall observing strategy (Guiglion et al., p. 17) to enable the ESO community to develop complementary surveys using the roughly 4.8/2.4 million LRS/HRS fibre-hours available to them in the first 5-year survey. The process of integrating community observing programmes into the 4MOST survey programme is described in the next section.

## Community programmes

In designing the 4MOST operations system, the aim has been to follow normal ESO operations as much as possible. This means that 4MOST follows the ESO Public Surveys sequence of programme selection (Phase 1), observation preparation (Phase 2), programme execution at the telescope, and finally data reduction, analysis, and publication (Phase 3). However, 4MOST, being a survey facility running typically many science programmes simultaneously in each observation, has required some modifications to the normal process, as described below.

As highlighted earlier, 4MOST Surveys have a duration of five years. This ensures that large projects can be accomplished with carefully crafted completeness goals and well understood selection functions. New programmes will be selected and started only once every five years and, after a short run-in period, the observing strategy will stay as stable as possible during each five-year survey programme. All surveys on 4MOST will be Public Surveys, which means that the raw data will be published immediately in the ESO archive and that the science teams of the surveys have an obligation to release higher-level data products that have legacy value for the community.

Table 2. 4MOST Consortium Surveys and their Principal Investigators.

| No | Survey Name | Survey (Co-)PI |
|---|---|---|
| S1 | Milky Way Halo LR Survey | Irwin (IoA), Helmi (RuG) |
| S2 | Milky Way Halo HR Survey | Christlieb (ZAH) |
| S3 | Milky Way Disc and Bulge LR Survey (4MIDABLE LR) | Chiappini, Minchev, Starkenburg (AIP) |
| S4 | Milky Way Disc and Bulge HR Survey (4MIDABLE HR) | Bensby (Lund), Bergemann (MPIA) |
| S5 | Galaxy Clusters Survey | Finoguenov (MPE) |
| S6 | AGN Survey | Merloni (MPE) |
| S7 | Galaxy Evolution Survey (WAVES) | Driver (UWA), Liske (UHH) |
| S8 | Cosmology Redshift Survey | Richard (CRAL), Kneib (EPFL) |
| S9 | Magellanic Clouds Survey (1001MC) | Cioni (AIP) |
| S10 | Time-Domain Extragalactic Survey (TiDES) | Sullivan (Southampton) |

The community can propose for one of two types of Survey programmes with 4MOST.

1) Participating Surveys from the ESO community will join the Consortium Surveys in a common observing programme, where they share the available fibres in each observing block and are "charged" fibre-hours only for their fraction of fibres used. They also share the time spent on any duplicate targets in common between surveys, get full access to all data from the Consortium and participating community programmes, and are invited to collaborate in the higher-level data analysis and publication efforts.

2) Non-Participating Surveys get their own (half) nights on the telescope and will be "charged" fibre-hours for the full 2436 fibres during that time regardless of whether they can all be filled. These surveys will receive calibrated and extracted spectra from the Consortium data management system, but will not have access to any data other than their own and they will be responsible for delivering higher-level data products to the ESO archive on their own. While many aspects are the same for Participating and Non-Participating Surveys, critical differences during the various execution phases of the Surveys are highlighted below.

### Phase 1

4MOST Phase 1 will begin with a Call for Letters of Intent. Each Letter of Intent is expected to set out: the science goals of the proposed survey; a description of its scope (for example, the number of targets and their distribution on the sky, the targets' luminosity range, the approximate number of fibre-hours needed); an initial list of Survey team members and their roles (i.e., a simple management plan); and whether the proposal is for a Participating or Non-Participating Survey. To estimate the feasibility and scope of the observations an Exposure Time Calculator (ETC) will be provided through an ESO web interface for single targets, and through an ETC tool from the 4MOST Consortium for many targets at once. After a peer review of the Letters of Intent that will be managed by ESO, a number of teams will be invited to respond to the 4MOST Call for Proposals, at which time ESO may suggest that some of the community proposals merge with other community or Consortium proposals.

At this stage a more detailed science case will be required as well as a full (mock) target catalogue with template spectra, spectral success criteria, and a total survey goal encapsulated by a figure of merit. A web-based version of the 4MOST Facility Simulator (4FS) will be provided, allowing proposers to check the feasibility of their proposed survey. 4FS will provide an estimate of the number of successfully observed targets in a five-year survey when run either standalone (Non-Participating proposals) or in conjunction with the Consortium Surveys (Participating proposals) and the required number of fibre-hours. Clearly, proposals that are well matched to the overall observing strategy of 4MOST as described in the 4MOST Survey Plan article in this edition (for example, surveys with sparsely distributed targets or with looser completeness requirements) have a higher chance of being successfully executed in the amount of time available.

After selection of all Consortium and Community Surveys through ESO's peer review process for Public Surveys proposals, the selected programmes will be invited to submit survey management plans, approval of which by the ESO Director General is mandatory before the final acceptance of a Public Survey. The survey management plan will contain a detailed list of science data products and timeline for their release. For Consortium and Participating Community Surveys a single, joint survey management plan will be delivered. For Non-Participating Community Surveys, each Survey PI will be responsible for the delivery of a survey management plan.

### Phase 2

After selection, the members of the Participating Community Surveys will join the Consortium Surveys to form the joint Science Team. The Community Survey PIs will become members of the Science Coordination Board and it is expected that the Community Surveys will provide staff effort to the different 4MOST working groups, most notably those on survey strategy, selection functions, quality assurance, and, if they so wish, higher-level pipelines. The target catalogues of the Community Surveys will be merged with those of the Consortium and through an iterative process a joint survey plan will be developed to observe all targets. Once the final observing strategy has been agreed upon, only small changes in strategy will be allowed during the operations phase without approval by the SCB and/or ESO. The 4MOST Operations Group provided by the Consortium will create all Observing Blocks running on 4MOST.

Non-Participating Surveys will not join the Science Team, but will be provided with software to create and submit their own Observing Blocks which will be scheduled on their assigned (half) nights. Any significant changes from the original Non-Participating Survey plan will have to be approved by ESO.

### Phase 3

As with ESO's Public Survey policies, 4MOST Survey programmes have data delivery obligations to ESO and its community. All 4MOST raw data will become available as soon as they have been ingested into the ESO archive at the end of each night. The raw data will be processed by the Consortium Data Management System to remove instrumental effects and create one-dimensional, flux- and wavelength-calibrated Level 1 (L1) spectra. The L1 data will be released yearly through the ESO archive. For Participating Surveys, dedicated classification, stellar and extragalactic pipelines run by Consortium working groups will produce Level 2 (L2) data products like object type likelihoods, stellar parameters, elemental abundances and redshifts, etc. These products will be released through the ESO archive on a schedule to be agreed upon with ESO before the start of the observations. All L1 and L2 products will also be released through the 4MOST World Archive operated by the Consortium, which will also contain matched catalogues from other facilities and added value catalogues with data processed beyond the standard pipelines. While the Consortium will take care of uploading the L1 and L2 products to the ESO archive for the joint Science Team, Non-Participating Surveys will have to produce and upload their own L2 products to ESO.

## Policies

Given the joint use of the available fibres and the corresponding mixed nature of the data products, members of the Consortium Surveys and Participating Community Surveys, i.e., members of the joint Science Team, have to abide by a number of policies to ensure fair use of data and a fair return on investment. Community Survey membership will be limited to those on the original proposal plus up to 15 additional members added at a later stage if a certain capability or expertise is needed that is not available within the Science Team. Participating Community Survey targets may overlap by a maximum of 20% with Consortium targets, but will share the required "cost" in exposure time for the overlap, allowing both surveys to do more in their allotted amount of fibre-hours. All data products are shared among all Science Team members. However, all science exploitation shall take place in projects announced to the whole Science Team and restrictions regarding this exploitation may be applied when a new project overlaps significantly with an existing PhD project or with the core science of a Survey that the project proposer is not a member of. Full details of these Science Team policies as approved by ESO will be released alongside a Code of Conduct when the Call for Letters of Intent is published. By submitting a Participating Survey programme the proposers implicitly agree to comply with these policies.

For Non-Participating Surveys there may be at most a 30% overlap in targets with other Surveys and they will not share exposure time with other Surveys. This means that any duplicate targets in Non-Participating Surveys will be observed twice as there is no means to coordinate the effort with other Surveys. Non-Participating Surveys are free to devise their own membership, data access, and publication policies.

## Further information

ESO and the 4MOST Consortium are jointly organising the "Preparing for 4MOST" workshop, which will take place at ESO Garching on 6–8 May 2019. The purpose of this workshop is to transfer knowledge from the 4MOST Consortium to the broader ESO community, and hence to prepare the community for the exciting scientific opportunity to use 4MOST. This will assist potential community PIs to successfully respond to the Call, and will foster scientific collabora-

Table 3. 4MOST Consortium institutes and their main roles in the Project.

<table>
<tr><th>Institute</th><th>Instrument responsibility</th><th>Science lead responsibility</th></tr>
<tr><td>Leibniz-Institut für Astrophysik Potsdam (AIP)</td><td>Management and system engineering, telescope interface (including WFC), metrology, fibre system, instrument control software, System AIV and commissioning</td><td>Milky Way Disc and Bulge LR Survey, Cosmology Redshift Survey, Magellanic Clouds Survey</td></tr>
<tr><td>Australian Astronomical Optics – Macquarie (AAO)</td><td>Fibre positioner</td><td>Galaxy Evolution Survey</td></tr>
<tr><td>Centre de Recherche Astrophysique de Lyon (CRAL)</td><td>Low-Resolution spectrographs</td><td>Cosmology Redshift Survey</td></tr>
<tr><td>European Southern Observatory (ESO)</td><td>Detectors system</td><td></td></tr>
<tr><td>Institute of Astronomy, Cambridge (IoA)</td><td>Data management system</td><td>Milky Way Halo LR Survey</td></tr>
<tr><td>Max-Planck-Institut für Astronomie (MPIA)</td><td>Instrument control system hardware</td><td>Milky Way Disc and Bulge HR Survey</td></tr>
<tr><td>Max-Planck-Institut für extraterrestrische Physik (MPE)</td><td>Science operations system</td><td>Galaxy Clusters Survey, AGN Survey</td></tr>
<tr><td>Zentrum für Astronomie der Universität Heidelberg (ZAH)</td><td>High-Resolution spectrograph, Instrument control system software</td><td>Milky Way Halo HR Survey</td></tr>
<tr><td>NOVA/ASTRON Dwingeloo</td><td>Calibration system</td><td></td></tr>
<tr><td>Rijksuniversiteit Groningen (RuG)</td><td></td><td>Milky Way Halo LR Survey</td></tr>
<tr><td>Lund University (Lund)</td><td></td><td rowspan="2">Milky Way Disc and Bulge HR Survey</td></tr>
<tr><td>Uppsala universitet (UU)</td><td></td></tr>
<tr><td>Universität Hamburg (UHH)</td><td></td><td rowspan="2">Galaxy Evolution Survey</td></tr>
<tr><td>University of Western Australia (UWA)</td><td></td></tr>
<tr><td>École polytechnique fédérale de Lausanne (EPFL)</td><td></td><td>Cosmology Redshift Survey</td></tr>
</table>

tions between the community and the 4MOST Consortium. Members of the community who are considering applying for a 4MOST Public Survey are strongly encouraged to attend this meeting in order to obtain detailed information, and to have the opportunity to ask questions, exchange ideas, and build collaborations.

The latest information about 4MOST and its planned surveys is available on its website[1]. Information can also be obtained through the 4MOST helpdesk, which can be reached through ESO's User Support Department[2], the 4MOST web site, or by mailing the project directly[3].

## Schedule

The 4MOST Project moved into full construction after passing Final Design Review-1 in May 2018. Major milestones in further development and construction are the release of the Call for Letters of Intent in the second half of 2019 which will have a submission deadline about 2 months later, completion of the system integration in Potsdam in July 2021, passing the full system test including operations rehearsals for the Preliminary Acceptance Europe by February 2022, and the installation and commissioning of the facility at VISTA for Provisional Acceptance Chile in November 2022, after which the first five year survey will start.

## Consortium and Minor Participants Institutes

The 4MOST Consortium institutes and their main roles in the project are listed in Table 3. The following Minor Participant institutes are also contributing to the development of 4MOST: Durham University, University of Sussex, University College London, Institute for Astrophysics Göttingen (IAG), University of Warwick, University of Hull, Universität Potsdam, Laboratoire d'Etudes des Galaxies, Etoiles, Physique et Instrumentation (GEPI), IN2P3/Laboratoire des Matériaux Avancés (L.M.A.); and for the TiDES Survey: Lancaster University, Queen's University Belfast, University of Portsmouth, and University of Southampton.

### Acknowledgements

Financial support for 4MOST from the Knut and Alice Wallenberg's Foundation, the German Federal Ministry of Education and Research (BMBF) via Verbundforschungs grants 05A14BA2, 05A17BA3 and 05A17VH4, and from the German Research Foundation (DFG) via Sonderforschungsbereich SFB 881 "The Milky Way System" is gratefully acknowledged.

### Links

[1] The 4MOST website: www.4most.eu
[2] ESO's User Support Helpdesk: usd-help@eso.org
[3] 4MOST project mailing address: help@4most.eu

### Notes

[a] Roelof de Jong is the 4MOST Principal Investigator.

G. Gillet/ESO

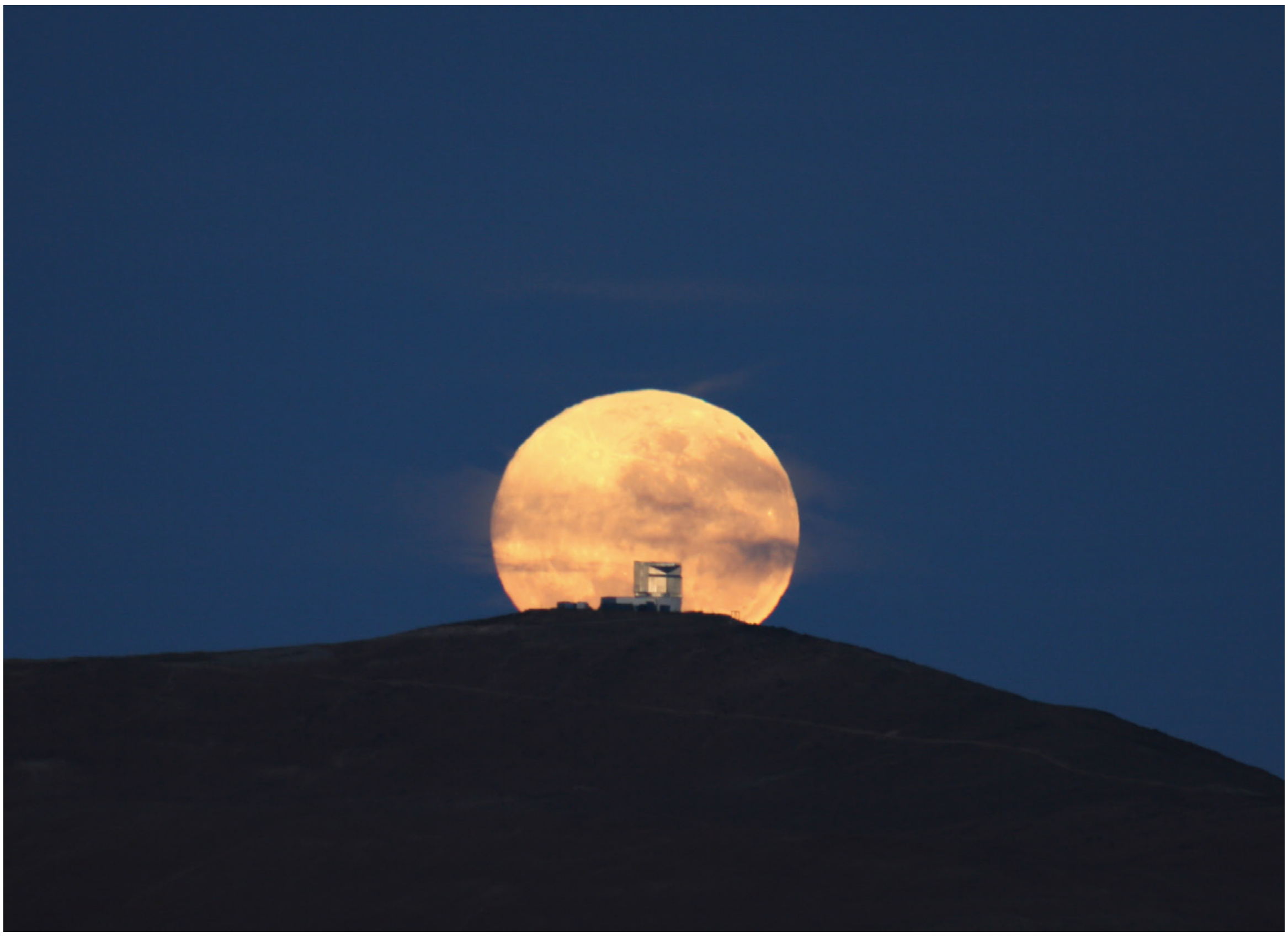

The full moon sets behind VISTA near Paranal.